\documentclass[preprint,12pt,a4paper]{elsarticle}
\usepackage{graphics}

\journal{Nuclear Physics A}

\begin{document}

\begin{frontmatter}

\title{Broad resonances and beta-decay}

\author[aar]{K. Riisager\corref{cor1}}
\author[aar]{H.O.U. Fynbo}
\author[aar]{S. Hyldegaard}
\author[aar]{A.S. Jensen}

\address[aar]{Department of Physics and Astronomy, Aarhus University,
 DK-8000 Aarhus C, Denmark}
\cortext[cor1]{Corresponding author, kvr@phys.au.dk}

%

\begin{abstract}
Beta-decay into broad resonances gives a distorted lineshape in the
observed energy spectrum. Part of the distortion arises from the
phase space factor, but we show that the beta-decay matrix element may
also contribute.
Based on a schematic model for p-wave continuum
neutron states it is argued that beta-decay directly to the continuum
should be considered as a possible contributing mechanism in many
decays close to the driplines.
The signatures in R-matrix fits for such decays
directly to continuum states are discussed and illustrated through
an analysis of the beta-decay of $^8$B into $2^+$ states in $^8$Be. 
\end{abstract}

\end{frontmatter}

\section{Introduction}
The concept of a resonance is pervasive in quantum physics as applied
e.g.\ on nuclear, particle, atomic and molecular phenomena. However, a
closer look at the literature shows that there is no unique way of
defining a resonance. When applied in data analysis, varying
definitions can give different results for broad resonances
\cite{Til02,PDG14,Boh05} or may even at some point become impossible
to apply.  There are two distinct aspects of a resonance, the first
being as a state of a (continuum) system in analogy with a bound
state, the second as characterizing an enhanced response to a
disturbance. We shall here mainly be concerned with the first aspect.

We shall focus in this paper on beta-delayed particle emission
processes that traditionally are considered to proceed through
(resonance) states in the beta-decay daughter and shall argue that
beta-decays directly to the continuum should be taken into account in
more situations than done so far. Before dealing specifically with
beta-decay we shall in section 2
remind the reader about some aspects of the
description of resonances and how well they perform when applied to
broad resonances. This will be illustrated in section 3
through simple model
calculations. One goal is to clarify whether parameters taken
from analyses of beta-decay data can be used in modelling of other
nuclear processes proceeding through the same ``reaction channels'',
we show in section 4 that loosely bound initial states may require special
considerations.  In the limit of very broad resonances it is not
possible to decouple the population of the resonance from its
decay. The properties of the resonance, its position and width (or
more generally its shape), will differ when populated via different
mechanisms.  Since analyses of decays through broad levels are often
made via R-matrix fits we shall discuss (section 4.3) 
what effects may occur there. Finally, section 5 discusses how our
results may be generalized and section 6 presents our conclusion.

It may be appropriate first to recall that a resonance as such does
not correspond directly to any physical observable. However, the
resonance concept can be very useful in describing the evolution of a
system, e.g.\ as a response to an external probe, namely by employing
resonances as basis states in the description. We are of course never
forced to use a specific set of basis states, but narrow resonances in
particular seem a natural choice. In the opposite limit of
very broad structures the alternative description in terms of ``pure''
continuum states (a basis defined by the asymptotic behaviour of the
wavefunctions) may seem the natural one. It is important to note that
both descriptions are valid and, at least for structures of
intermediate width, can be used in practice. A practical example can
be found in the calculation \cite{Myo98} of the dipole strength
function for $^{11}$Be where a basis combining resonances and
continuum contributions is used and it is demonstrated explicitly how
the number of included resonances can be varied without changing the
result. (The complex scaling method \cite{Myo14} used in this work can
be related to the Berggren decomposition of the continuum
\cite{Ber68,Ber93}.)

From such general considerations it appears that it is to some extent a matter
of convenience whether one interprets an experimental spectrum in
terms of resonances or not. This point has been made very clearly by
Dalitz \cite{Dal70,Alb84}. The two possible descriptions, emphasizing
resonances or continuum states, may be thought of as complementary,
and the question of whether a process happens resonantly or not will
as noted earlier \cite{Boh69,Bre64}
not always have a unique, or meaningful, answer. Nevertheless it is
worthwhile to explore how far the resonance concept may be taken, to
see how different resonance definitions relate to experimental
observables, and to determine when corrections must be included.


\section{Limits for resonant behaviour}
Most resonance definitions are conceptual or formal, but do agree in
the case of narrow resonances. Resonances may also be identified in
experimental spectra, and a standard requirement has been that several
different observables should give consistent (energy and width)
parameters for a given resonance, the point being that resonances
should be an intrinsic property of the system studied and ideally
not influenced by the ways of exciting it. For increasingly
broader resonances the deduced parameters can no longer be expected to
be identical for the different conceptual definitions or different
observables.

Blatt and Weisskopf \cite{Bla52} start their exposition of nuclear
resonances by considering the relative amplitude of wavefunctions
inside and outside the nucleus. A resonance corresponds to an
enhancement of the interior wavefunction and thereby to a spatial
concentration. Another possibility is to look at the time delay
between incoming and outgoing wave packets that should be large, see eg.\
\cite{Dal70,Tay72}.  That these two criteria are equivalent is shown
by the discussion in \cite{Fon78} on how the presence of an unstable
state leads to localization. As $\Gamma$ (the resonance width
parameter) increases these two ways of identifying resonances become
less clear: the enhancement of the interior wavefunction for a given
energy will decrease and the delay time (lifetime) will approach the
transit time of the constituents across the nucleus.

It is well known that a state decaying exponentially in time will have an
energy distribution given by a Breit-Wigner shape, but this
distribution can also be derived in several other ways, see
e.g. \cite{Bla52,Tay72,Fon78,Bre59}.  For broad levels
the pure Breit-Wigner shape has to be corrected (energy distributions
do not extend below zero energy, nor to arbitrarily high
energies). Furthermore it becomes harder to unambiguously interpret
broad structures in energy spectra as (distorted) Breit-Wigner
distributions since bumps in an energy distribution may occur for a
number of other reasons \cite{Dal70,Alb84}.

When following the evolution of a bound state as its energy is
increased and crosses the threshold to the continuum, it is tempting
to extend the eigenvalue description by employing complex energies and
letting the state acquire an imaginary energy corresponding to half its
width. One often in theoretical descriptions starts from complex
eigenstates with purely outgoing waves, the Gamow states
\cite{Civ04}. This naturally leads to the description of resonances as
poles (at complex values) in the S-matrix \cite{Tay72,Sim78}. This is
the definition employed in most, but not all \cite{Ort90}, treatments
in mathematical physics. It leads to very elegant formulations, but
also runs into problems for very broad states. On one hand, as pointed
out by Sitenko \cite{Sit71}, the assumption of purely outgoing waves
is no longer realistic when the time delay is very short. 
On the other hand, as noted e.g.\ in \cite{Tay72}, we can easily
identify poles close to the real axis in experimental data (that are
on the real axis), but as widths become larger and the poles move away
from the real axis it becomes increasingly harder to make this
identification.  In practical situations one often has to introduce
more than one state and thereby many parameters for the states to
determine from the experimental spectrum, examples will be given in
section \ref{sec:Rmat}. In the extreme situation there is ``more input
from the theoretical skeleton than from the experimental data''.

See also \cite{Alb84,Bad82,Ela87,Boh93,Moi98,Fer02,Oko03,Hat10,Car14} for more (and
complementary) accounts of the resonance phenomenon.

\section{Schematic model}
The changes that occur as resonances become broader will be
illustrated in a schematic model where the many nuclear degrees of
freedom are restricted to the motion of a single neutron. The
corresponding strong
potential is taken to be a square well with radius $a=4$ fm and depth
$V_0>0$.  For a positive energy $E=(\hbar k)^2/2\mu$ (for definiteness
we shall take the reduced mass $\mu$ to be 10/11 of the neutron mass)
and an angular momentum $l$ the radial wavefunction, $R(r)$, inside the
potential is proportional to $j_l(Kr)$ where $E+V_0 = (\hbar
K)^2/2\mu$. Outside the potential the radial wavefunction is
normalized to be the
linear combination $[\cos \delta_l j_l(kr) - \sin\delta_l y_l(kr)]k$
of the regular and irregular spherical Bessel function of order
$l$. Here $\delta_l$ is the phase shift that is found by matching the
inner and outer wavefunction.  The explicit factor of $k$ implies that
the wavefunction is normalized to approach
$\sin(kr+\delta_l-l\pi/2)/r$ for large radii $r$.
We shall mainly consider p-waves, this is the simplest case where
resonances occur since s-waves that have no confining potential form
virtual states instead. 

\begin{figure}[thb]
\centering
     \includegraphics[width=10.cm,clip]{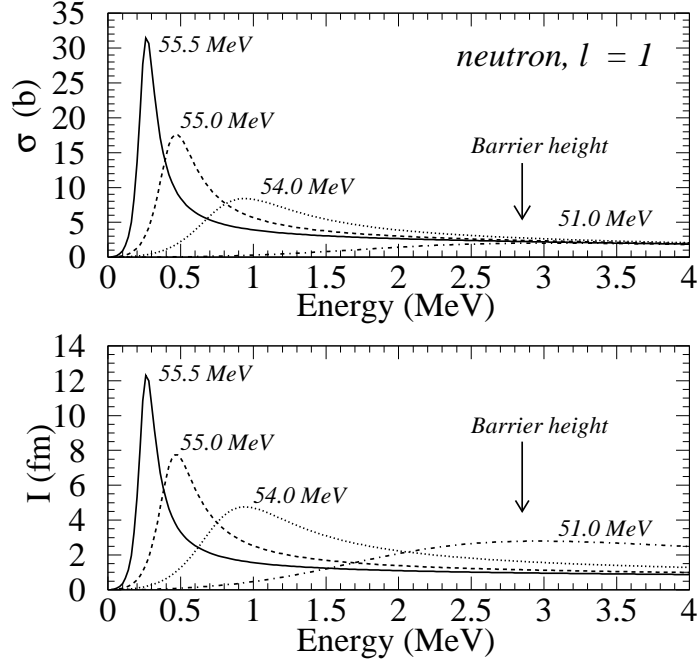} 
     \caption{Upper panel: Elastic cross-section, lower panel: the
       integral of the squared wavefunction inside the potential. Both
       are calculated for p-wave neutrons in a square well of radius
       4.0 fm and shown as a function of energy for different well depths
       as marked in the figure. The height of the angular momentum
       barrier is marked by an arrow.}
\label{fig:int_e1} 
\end{figure}

\subsection{Elastic scattering}
This simple model will now be used to compare different resonance
definitions. First of all, a phase shift that goes through $\pi/2$ is
often used as a criterion. This is closely related to elastic
scattering for which the cross section is
\begin{equation}
  \sigma_l = \frac{\pi}{k^2} (2l+1)|e^{2i\delta_l}-1|^2 \;.
\end{equation}
The factor $|e^{2i\delta_l}-1| = 2\sin \delta_l$ is clearly maximal when the phase
shift equals $\pi/2+n\pi$. However, for broad levels the variation of
the front factor $k^{-2}$ shifts down the maximum for the cross
section to lower energies.  As a
second way to define resonances, one may look at the enhancement of the
interior wavefunction that can be quantified through the integral $I =
\int_o^a |R(r)|^2r^2\mathrm{d}r$ of the squared wavefunction over the
potential range. If no potential is present, i.e.\ the neutron is in
an s-wave and $V_0 = 0$ MeV, the integral averages to $a/2 = 2$ fm.

Figure \ref{fig:int_e1} shows the calculated elastic cross-section
$\sigma$ and the integral $I$ of the interior wavefunction as a
function of energy for different choices of the potential depth $V_0$
just above 50 MeV (in this range there will be one bound state in the
potential).  As seen in the figure peaks occur as long as they are
well below the height of the potential barrier. The peak shape of both
$\sigma$ and $I$ become wide and asymmetric as the peak positions
increase, note that the peak positions and widths are quite similar in
the two cases. There is still a broad structure visible in $I$ for
$V_0 = 51.0$ MeV, but a less clear signal in $\sigma$. Furthermore,
the positions where the phase shift goes through $\pi/2$ deviates more
and more from the visual peak and the phase shift never reaches
$\pi/2$ for potential depth 51.0 MeV. The resonance positions according
to these two definitions are given in table \ref{tab:Eres} in column
two and four. The peak position for the integral $I$ is essentially
identical to the position of the maximum of the elastic scattering cross
section. The lower part of table \ref{tab:Eres} have potential depths
corresponding to wavefunctions with no nodes inside the potential, for
the upper part there is one node.

%

\subsection{A second look at resonance definitions} \label{sec:2res}
The numerical results shown above suggests that resonances should be
characterized by being localized in space as well as peaked in energy, this
seems natural for the case of short-ranged potentials considered in
this paper. The implications of this point of view will be pursued in
the following sections, but a few observations can be made already.

\begin{table}
\centering
\caption{The position of a resonance for p-wave neutrons in a square
  well of radius 4 fm and depth $V_0$ according to four different
  definitions: the total or resonant phase shift, $\delta_l$ or
  $\delta_R$, going through
  $\pi/2$ and the maxima of the elastic cross-section or the
  beta-strength distribution. All energies are in units of MeV.}
\label{tab:Eres} 
\begin{tabular}{c|cccc}
$V_0$  & \multicolumn{4}{c}{Resonance energy}  \\
         & $\delta_l = \frac{\pi}{2}$ & $\delta_R = \frac{\pi}{2}$ & max
$\sigma_l$ & max $B(E)_{E_i = 1 \mathrm{MeV}}$ \\ \hline
55.5  & 0.278 & 0.276 & 0.266 & 0.256 \\
55.0  & 0.510 & 0.497 & 0.469 & 0.436 \\
54.5  & 0.791 & 0.748 & 0.693 & 0.616 \\
54.0  & 1.146 & 1.028 & 0.942 & 0.797 \\
53.0  & 2.358 & 1.672 & 1.518 & 1.154 \\
52.0  &   --    & 2.411 & 2.216 & 1.500 \\
51.0  &   --    & 3.219 & 3.05   & 1.831 \\ \hline
12.0  & 0.975 & 0.894 & 0.814 & 0.705 \\
11.5  & 1.379 & 1.177 & 1.055 & 0.868 \\
11.0  & 1.970 & 1.486 & 1.314 & 1.027 \\ \hline
\end{tabular}
\end{table}

The enhancement of the wavefunction inside the potential will also be
present in the region just outside. The integral $I^{out} =
\int_a^{2a} |R(r)|^2r^2\mathrm{d}r$ will in most cases be larger than
$I$ for states of energy close to the peak energy. This will be
important when we shall consider transition matrix elements in the
next section, but it also underlines that resonance wavefunctions do
differ qualitatively from bound state wavefunctions.  An exception to
this is halo states that also extend significantly beyond the
potential range.

The wavefunction inside the potential, $j_l(Kr)$ in our model, changes
explicitly with energy through $K$. The relative change across a peak
of width $\Gamma$
is easily shown to be $\Delta K/K = \Gamma/2/(E+V_0)$ and can become
appreciable for very broad states. In this limit both the
normalization and the functional shape of the interior resonance
wavefunction is no longer unique. This situation can be contrasted to
the one encountered in R-matrix theory that is often used in advanced
analyses of experimental data. 
We refer to \cite{Bre59,Lan58,Bay10} for a full account of R-matrix
theory and only summarize the main features here:
Internal levels $\lambda$ appear
from quantization within the channel radius $a_c$, they have energies
$E_{\lambda}$ and amplitude $\gamma_{\lambda}$ for coupling to
a decay channel. The internal wavefunctions form a discrete
basis set for the interior, but the levels cannot automatically be
identified as resonances, we shall elaborate on this in section \ref{sec:Rmat}. For
the cases where such an identification can be made Lane and Thomas showed
in section XII.2 of their review \cite{Lan58} that for
an internal wavefunction $\Psi$ formed with unit incoming flux
(spherical wave) one has at a resonance for the integral over the
internal region:
\begin{equation}
  \int |\Psi|^2 \mathrm{d}^3r = \frac{\hbar \Gamma_{\lambda}}
     {(E_{\lambda} +\Delta -E)^2+\Gamma_{\lambda}^2/4} \;,
\end{equation}
where $\Gamma_{\lambda} = 2 P \gamma_{\lambda}^2$, $\Delta =
-(S-B)\gamma_{\lambda}^2$, the penetrability and shift factor for the
channel, $P$ and $S$, are energy dependent and the boundary parameter $B$
normally is set equal to $S$ at resonance. This explicitly shows how
the single-level (Breit-Wigner) resonance formula is related to our
integral $I$. If one here approximates
$S$ by a linear function (section XII.3 of \cite{Lan58}) one obtains
an expression with ``observed'' parameters rather than the above, more
formal R-matrix parameters:
\begin{equation}
   C \frac{\hbar \Gamma_0}{(E_0 -E)^2+\Gamma_0^2/4} \;\;, \;
  C = \frac{1}{1+\gamma_{\lambda}^2 \mathrm{d}S/\mathrm{d}E}\; ,
\end{equation}
where now $\Gamma_0 = C\Gamma_{\lambda}$ corresponding to a
renormalization of $\gamma_{\lambda}^2$ and $E_0 =
E_{\lambda}+\Delta$. However, the energy dependence of the
penetrability implies that the maximum of the distribution may be
shifted away from $E_0$ \cite{Bar64}. In the limit of small shifts one
can for neutrons derive that the maximum is positioned at
\begin{equation}  \label{eq:eshift}
  E_{max} = E_0 \left[ 1- \alpha \left( \frac{\Gamma_0}{4 E_0} \right)^2
  \right] \;,
\end{equation}
where the parameter $\alpha$ for a given angular momentum $l$
increases from 1 far above the barrier to $2l+1$ well below the
barrier. For p-waves $\alpha = [3+(ka)^2]/[1+(ka)^2]$.

For an isolated resonance the phase shift in
R-matrix theory consists of two contributions $\delta_l =
\phi_l+\delta_R$. Here $\phi_l = \tan^{-1}(j_l/y_l)$ is the hard
sphere phase shift and 
\begin{equation}  \label{eq:phres}
   \delta_R = \tan^{-1}\left(
     \frac{\Gamma_{\lambda}/2}{E_{\lambda}+\Delta-E} \right)
\end{equation}
is the
contribution from the resonance. Since the resonance position in
R-matrix theory is
defined by $E_{\lambda}+\Delta-E = 0$ it corresponds to $\delta_R$
being $\pi/2$. The corresponding resonance positions have been
evaluated and are also given as the third column in table
\ref{tab:Eres}. (The final column will be explained in the next
section.) Using $\delta_R$ rather than $\delta_l$ clearly gives more
appropriate resonance positions; the increase of more than one MeV in
$\delta_l$-resonance position as $V_0$ is decreased from 54 MeV to 53
MeV is not sensible. For broad resonances $\delta_R$ gives a resonance
position above the peak position, but this result actually depends on
the value chosen for the channel radius $a_c$ used to evaluate
$\phi_l$. It was here taken equal to $a = 4$ fm, but increasing it to
6 fm will move the extracted resonance positions down e.g.\ to 0.477
MeV and 1.380 MeV for $V_0 = $ 55 MeV and 53 MeV. This highlights one
of the conceptual problems in R-matrix theory, namely the possible
dependence of results on the channel radius, see e.g.\ the discussion
in \cite{Bay10}. We shall return to the
question of resonance positions extracted from R-matrix theory in
section \ref{sec:Rmat}.

\section{Beta-delayed neutron emission}
Beta-decay may populate broad resonances,
in particular in light nuclei,
and could conceptually present a new angle on the issue since one here
enters ``abruptly'' into a strongly interacting system, thereby
circumventing Sitenko's point presented above. Broad resonances could
therefore appear differently when populated in beta decay rather than
in nuclear reactions. Population through gamma decays (or
photo-dissociation) will be similar, but with a slightly more complex
operator. Although there are fewer detailed studies of gamma decays to
broad levels, some exist \cite{DeB95,Kir09,Dat13} and related studies
on photo-dissociation processes can also be relevant.  References to
the extensive literature on beta decay to continuum levels may be found
e.g.\ through a recent review paper \cite{Pfu12}.

\subsection{Model calculations}
The operator for allowed beta-decay transforms a neutron into a
proton without further changes to the wavefunction. This is
approximated in the schematic model by assuming that a core nucleon is
transformed. The neutron is then involved through the overlap matrix
element between an initial state bound p-wave neutron (with a binding
energy $E_i$ that can be varied) and the above final state continuum
p-wave function. With our normalization of the latter wavefunction
the matrix element has dimension of a length, and the phase space
factor corresponding to the density of final neutron states is
$2/(\pi\hbar v)$ where $v = \hbar k/\mu$ is the velocity
(the factor can be derived from the more detailed
expressions in \cite{Bon88}). We follow \cite{Rii14} and define the
differential beta-strength $B(E)$ as the product of the matrix element
squared and the neutron phase space factor. The beta-decay rate,
$w(E)$, at a given energy $E$ is then proportional to $f(Q-E)B(E)$
where $f$ is the usual beta-decay phase space factor. The $f$-factor
has a substantial energy dependence which will move the observed peak
position by roughly $-(5/8)\Gamma_0^2/(Q-E_0)$, see \cite{Rii14} for
details. We shall here focus on the behaviour of $B(E)$ and note that
the $f$-factor can be divided out from experimental data so that its
effects can be removed.  We neglect for the moment any effects of
non-perfect overlap in the core, but return below to a more detailed
discussion of the assumptions made.

\begin{figure}[thb]
\centering
     \includegraphics[width=10.cm,clip]{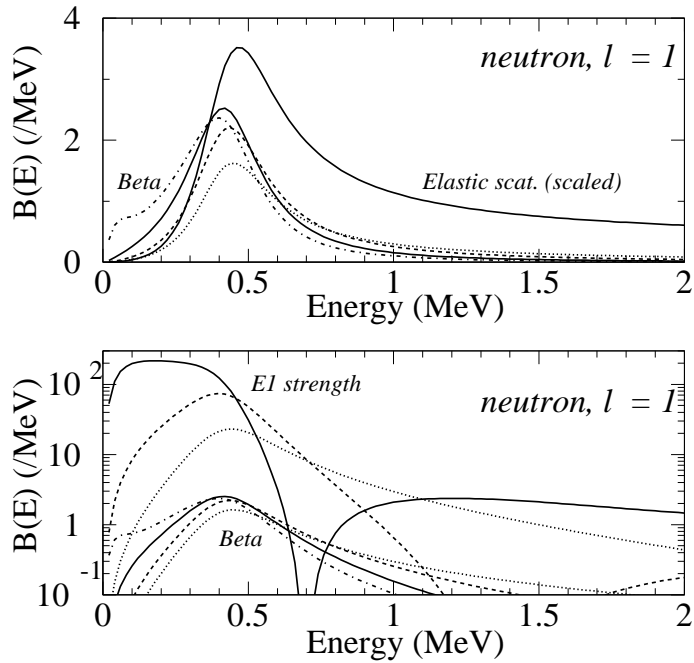} 
     \caption{The beta-strength to a p-wave neutron continuum for a
       square well radius 4.0 fm and potential depth 55.0 MeV is shown
       as a function of energy. Initial
       state binding energy is 5.0 MeV (dotted line), 1.0 MeV (dashed
       line), 0.2 MeV (solid line) and 0.02 MeV (dash-dotted line),
       respectively. Upper panel also shows the elastic cross-section
       (in barn) scaled down by a factor 5. Lower panel also shows the
       E1 strength for initial s-wave states with binding energies
       5.0 MeV, 1.0 MeV and 0.2 MeV.}
\label{fig:beta2} 
\end{figure}

For initial neutron states that are well bound so that the
wavefunction is mainly inside the potential one could expect the
square of the overlap matrix element to be similar to the integral $I$
from the previous section.  However, the beta-strength contains an
extra factor $1/v$ that will move the peak position down and reduce
its width. This can be seen in the upper panel of figure
\ref{fig:beta2} by comparing the (rescaled) elastic cross-section with
the differential beta-strength for a 5 MeV initial state (dotted
line). For initial states with smaller binding energy the wavefunction
will extend beyond the potential and there will be a contribution to
the overlap matrix element also from the external region. This leads
to a further shift of the peak position and modification of the line
shape as shown in figure \ref{fig:beta2}.  The area under the curves
is the same (due to the beta-decay strength sum rule), and the strength
is moved from higher energies to lower energies thereby shifting the
peak downwards, for the very low initial binding energy of 20 keV a
low-energy shoulder develops. These changes may be
understood as follows. Less initial binding energy $E_i$ gives a wavefunction with
slower radial fall-off which increases the importance of the external
contribution. This enhances the overlap for final state energies $E$
of order and smaller than $E_i$, but when $E$ increases the
oscillations in the final state wavefunction become more rapid and the
overlap is suppressed. It must be stressed that all changes seen in
figure \ref{fig:beta2} for different $E_i$ are due to the structure of
the initial state. The final state continuum is exactly the same in
all calculations.

The lower panel of the figure also shows schematic results
corresponding to ($\gamma$,n) dissociation reactions with E1
transitions. The initial states are here taken as neutron s-waves of
different energy, and the matrix element now includes a factor $r$ as
appropriate for the E1 operator. The quantity plotted is the matrix
element squared times $2/(\pi\hbar v)$. The extra factor of $r$ in the
matrix elements (and the fact that the initial s-wave neutron state
can be more extended than a p-wave) enhances the effects mentioned
above, but the overall qualitative behaviour is similar. One can
include the external electromagnetic contributions explicitly in a
resonance framework such as R-matrix theory (see section XIII.3 in
\cite{Lan58}), as also extended by Barker to the case of
beta-decays\footnote{Barker pointed out that this may be equivalent to
  considering processes with inverse time ordering \protect\cite{Gri60,Bor93},
  where the particle emission happens before the beta decay.}
\cite{Bar94}, but for the radiative capture case ((n,$\gamma$), the
inverse reaction to dissociation) it is customary to speak in terms of
a direct capture process, see e.g. \cite{Lan60}. For the E1 results in
figure \ref{fig:beta2} we would therefore speak of a transition from a
resonance dominated to a direct transition dominated process as $E_i$
is reduced. With the similar, but less pronounced, tendencies in beta
decay we may in a similar way attribute the changes to an increasing
contribution from decays directly into the continuum as $E_i$ is
decreased.

It is worth stressing that we are dealing with two limits of the same
physical process rather than two distinct reaction mechanisms. For the
case of E1 transitions the direct and resonant dominated limits
clearly correspond to the main contributions coming from external
distances and internal distances, respectively. It does not make much
sense to attempt to define a strict borderline between the two limits,
to give just one example a resonance may give clear interference effects
for processes that are mainly due to direct contributions. Before
leaving the electromagnetic processes it should be noted that the
drastic effects seen here are at least partly due to the initial
s-states. At low binding energies these are halo states, see e.g.\
\cite{Jen04}, and it is well established \cite{Ots94,Aum13} that their
pronounced low-energy E1-strength is non-resonant.

The maxima of the $B(E)$ distributions are also given in table
\ref{tab:Eres} for the case of initial binding energy $E_i$ of 1
MeV. The shift down in energy with respect to the maximum of the
elastic cross-section is sizeable in all cases, although always below
a ``half width at half maximum''.  The resonances appearing in our
model calculation have single-particle strength by construction. One
can reduce the width of the resonances by introducing an extra
delta-shell potential. We have checked that doing this reduces the
overall scale of the energy shifts, but peak shifts are still present
with a similar magnitude of the peak shift to peak width ratio.  If on
the other hand the beta strength is significantly smaller than unity
(the single-particle value) the beta transition may be due to other,
and smaller, components in the initial wavefunction where the
effective binding energies are larger. In this case the dependence on
$E_i$ will be strongly reduced.

\begin{figure}[thb]
\centering
     \includegraphics[width=10.cm,clip]{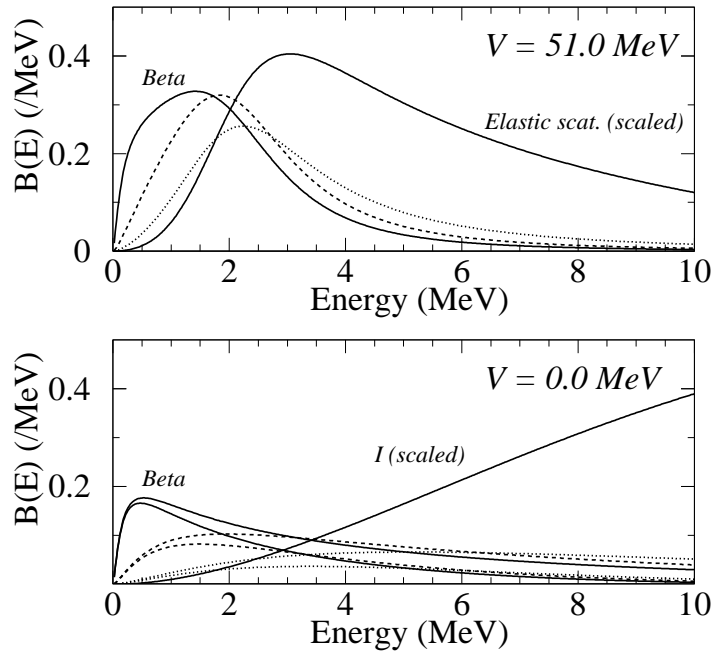} 
     \caption{The beta-strength to a p-wave neutron continuum for a
       square well radius 4.0 fm and potential depth 51.0 MeV (upper
       panel) or 0.0 MeV (lower panel). Initial state binding energy
       is 5.0 MeV (dotted line), 1.0 MeV (dashed line) and 0.2 MeV (solid
       line), respectively.
       Upper panel also shows the elastic cross-section
       (in barn) scaled down by a factor 5. Lower panel also shows the
       integral $I$ of the squared wavefunction over the interior
       scaled down by a factor 4. The two sets of beta-strength curves
       in the lower panel are for initial wavefunctions with no (upper
       curves) and one (lower curves) node inside the potential.}
\label{fig:beta3} 
\end{figure}

To further illustrate the effects that may occur in beta decay, two
extreme cases are shown in figure \ref{fig:beta3}, namely the final
state potential of 51.0 MeV that in figure \ref{fig:int_e1} gave broad
structure around the barrier height, and the case where the final
state potential vanishes. In the former case the effect of the initial
state binding energy is very high, shifting the peak position by more
than 1 MeV and giving significantly narrower distributions.  In the
latter case there are no resonances and the beta strength can only be
attributed to direct transitions to the continuum.  More than 50\% of
the beta strength sum rule value appears below 10 MeV for all shown
cases with no nodes in the initial wavefunction.

\begin{figure}[thb]
\centering
     \includegraphics[width=10.cm,clip]{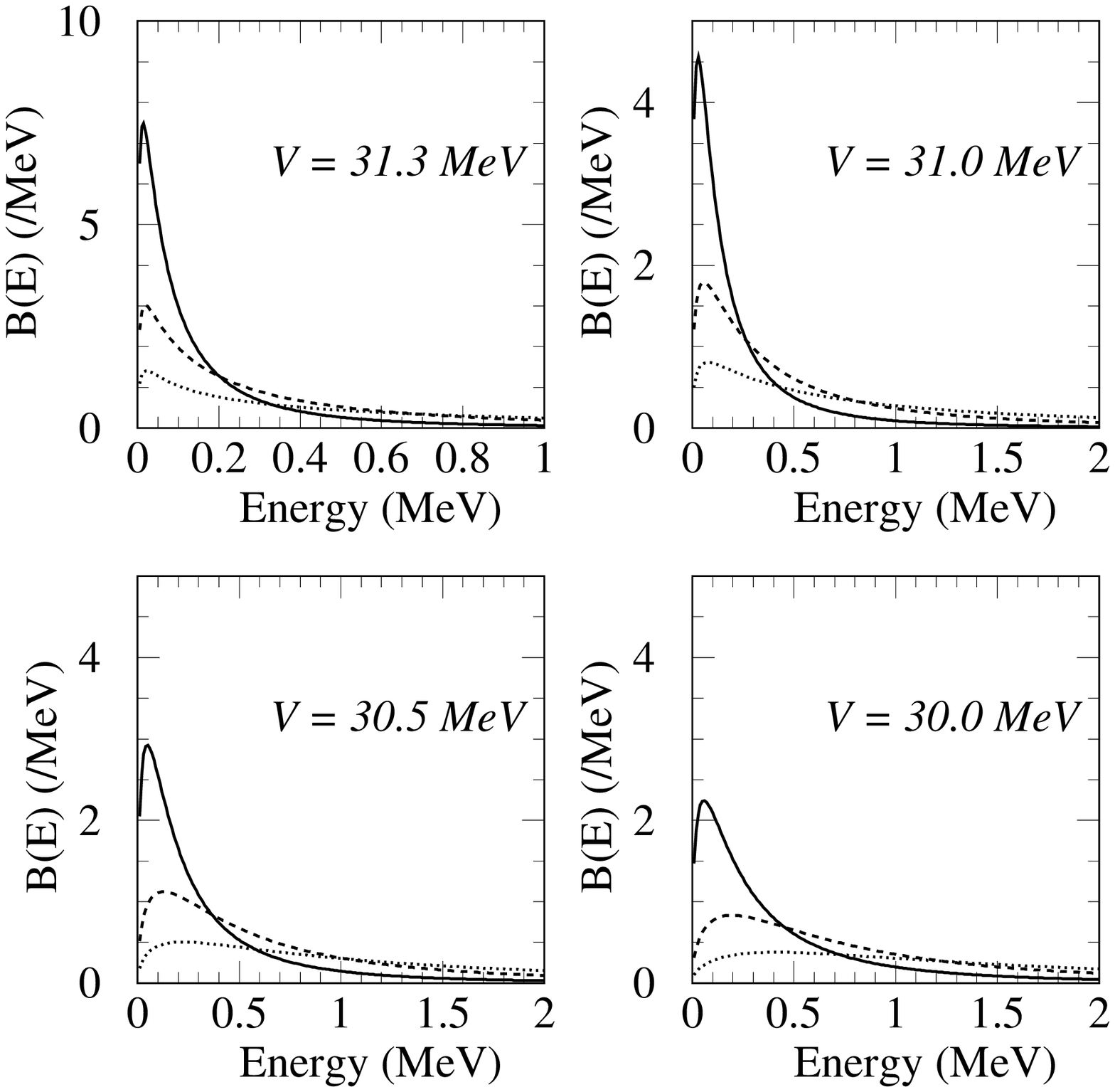} 
     \caption{The beta-strength to an s-wave neutron continuum for a
       square well radius 4.0 fm and potential depths as marked in the
       four panels. The corresponding scattering lengths are 30 fm, 14
       fm, 6.1 fm and 3.0 fm, respectively. The initial state binding energy
       is 5.0 MeV (dotted line), 1.0 MeV (dashed line) and 0.2 MeV (solid
       line), respectively. Note that the upper left panel has
       different ranges than the other panels.}
\label{fig:sbeta} 
\end{figure}

Since beta-decay gives resonance-like shapes even in rather extreme
situations, it is of interest to see also the behaviour for an s-wave
neutron continuum. Here the elastic scattering has a cross-section
that decreases monotonically, one talks of a virtual state rather than
a resonance and characterizes it by a scattering length $a_s$. The four
panels in figure \ref{fig:sbeta} show results for beta-decays into
such an s-wave continuum with scattering lengths decreasing from 30 fm
to 3 fm (the corresponding energy scales $\hbar^2/(2\mu a_s^2)$
go from 2.5 keV to 2.5 MeV).
There is now a distinct difference in the beta-strength distributions
obtained for different initial state binding energies, the figure
gives results from 0.2 MeV (a clear halo state) to 5 MeV (close to
standard binding energy). Still, all distributions peak at low energy (at a position
that depends as much on the scattering length as on the initial
binding energy) and have a pronounced asymmetric shape with a long
tail towards high energies. The tail is less extended for smaller
initial binding energies since the spatially larger initial
wavefunction gives rise to a quicker cancellation in the overlap
matrix element.  Most of the beta strength (for 1 MeV initial binding
energy more than 98\% of the sum rule) is lying between 0 MeV and 10
MeV. The shape of the distributions is not really resonance-like, but
varies significantly less than the elastic scattering cross-sections.

In summary, beta-decay may clearly distort the shape of a resonance.
The dependence of the beta-strength on the initial state is a
threshold effect, occuring when the initial binding energy is so low
that wavefunctions extend in a significant manner outside of the
nuclear core. It affects both peak position and peak shape and appears
even for an s-wave continuum. 
The binding energy dependence may be reduced for
transitions with low beta strength, but the distortion with respect
to elastic scattering should remain for very broad resonances.
It is interesting to compare this with
the detailed calculations of continuum-continuum E2 transitions in
$^8$Be discussed in \cite{Gar13}. In that case contributions to the
transition matrix element were found to arise mainly from the short
distance region. One may expect net contributions from larger
distances to appear only for continuum-continuum transitions where
both initial and final state have small $k$-values.

\subsection{Applicability of the model}
In the calculations above the beta-decay takes place in the core
whereas the initial neutron bound in a p-wave goes into a continuum
p-wave state. The model is clearly schematic and may exaggerate the
effect in that it has both maximum beta decay strength and full
single-particle strength for the subsequent particle emission.  The
interesting effects happen at low neutron binding and we are therefore
dealing with $\beta^-$ decays. To have a final state with good isospin
we need to add a component where the core is unchanged and the p-wave
neutron is transformed into a proton. In the usual core plus
single-particle model (see e.g.\ section 3-1 in \cite{Boh69}) the
component that we consider here with an unchanged neutron is the main
one in the isobaric analogue state (IAS) and the minor one in the
state of lower isospin. We shall consider the applicability of the
model separately for Fermi and Gamow-Teller transitions, but note that
the model could be appropriate for M1 excitations where the spin part
is dominating.

Standard $\beta^-$ decay does not populate the IAS (except for the
neutron and triton), but neutrino scattering $(\nu_e,e)$ may do so in
a Fermi transition. Still, the Coulomb energy keeps the neutron
threshold with the appropriate isospin above the IAS so that it only
decays by neutron emission through isospin impurities. To get a Fermi
contribution to beta-delayed neutron emission two conditions must be
met. First, the spatial overlap between the initial state and the IAS
must be less than one. This will happen if the initial and final
potentials differ slightly so that the two states have different
binding energy measured from their respective
thresholds. Experimentally such differences can be of order 100 keV;
our simple model in such cases gives feeding to the continuum of order
one percent for binding energies less than 1 MeV. (With exactly the
same potential in initial and final state there is of course no
continuum feeding.) The second condition is that isospin conservation
is broken. Large effects can only be expected for the low-lying
continuum where the asymmetry of the Coulomb effects is most
noticable. In the schematic model decays only take place via the
component where the core is beta-decaying and the requirement is
therefore that the core final state is unbound. This can be achieved
when the starting configuration is a two-neutron halo, $^{11}$Li being
the classic example.  A more detailed investigation of how large
fraction of the Fermi strength could realistically reside in the
low-lying continuum would be very interesting, the current model is
clearly not applicable but suggests that the one percent level is not
out of reach.

For Gamow-Teller transitions the beta strength will be reduced due to
the fact that the component we consider now is a small part of a
transition, but this is partly compensated by the factor 3 from the
spin operator.  The model may therefore be thought of as a schematic model
for beta-delayed neutron emission from nuclei with the last filled
neutrons in p-orbits. An interesting case is the beta decay of
$^{14}$Be where the largest branch goes to a state 
\cite{Bel97,Jep02,Aoi02} fed with a Gamow-Teller
strength close to 1 and situated about 300 keV above the one neutron
threshold in $^{14}$B. The beta-decay experiments report a neutron
line of energy 288(1) keV (corresponding to a level 308 keV above
threshold) and width 49(2) keV, but a somewhat asymmetric line
shape. A later $^{14}$Be(p,n)$^{14}$B reaction experiment at 69 MeV
\cite{Sat11} reported a transition corresponding to a level 304(4) keV
above threshold and a substantially larger width of 160(20) keV.
Even without detailed calculations of the charge exchange reaction
mechanism the difference in width seems too large to be accounted for
in our model: for a potential depth of 55.4 MeV elastic scattering
will have a peak position at 305 keV and a FWHM of 204 keV (a
single-particle strength), while for
an initial binding energy in the range 0.5-1.0 MeV the beta strength
will have a peak position around 15 keV lower and a FWHM around 170
keV. Nevertheless, a more detailed investigation of this case would be
very interesting.

Two possible cases occur for the neutron-rich boron isotopes, although
in both cases for transitions with beta strength of order 0.1 or less.
The nucleus $^{13}$B decays into a 1/2$^-$ 8.860(20) MeV level in
$^{13}$C with width 150(30) keV \cite{Ajz91}; here the latest
published experiment \cite{Alb74} is from 1974 and did not give
accurate energies for the position of the level.  A recent experiment
\cite{Uen13} on the beta decay of $^{17}$B found indications for a
2.5(7) MeV wide 5.04(2) MeV neutron line tentatively attributed to
neutron emission from a level at 6.08 MeV in $^{17}$C. If this is
confirmed one can in any case expect large effects for a resonance that is so
wide.
Other relevant examples may be the decays of nuclei such as $^8$He,
$^9$Li or heavier nuclei around $^{50}$K.

Similar effects to the ones exposed here may occur also in other
beta-delayed particle emission processes. 
Before discussing that it is
useful to look at how R-matrix fits behave in our model.

\subsection{R-matrix fits} \label{sec:Rmat} R-matrix theory allows to
include effects such as the penetrability and interference that can
alter the spectral shape for broad resonances. It is therefore often
employed in fits of experimental data in order to extract parameter
values.  We shall test here what happens for our model calculations
for the case of $V_0 = 55$ MeV. For the phase shifts equation
(\ref{eq:phres}) gives an almost perfect fit to the data when the hard
sphere phase shift is evaluated and the fit performed for $a_c = a =
4$ fm. The resonance position is at 0.497 MeV and the
$\gamma_{\lambda}$ parameter corresponds to a total level width equal
to a single particle unit (the Wigner limit). If $a_c$ is chosen as 6
fm (and the hard sphere phase shift corrected accordingly) it is not
possible to produce an acceptable fit with one resonance, to get this
one needs to introduce a second level or add a constant background
term to the R-matrix.

We have furthermore fitted the calculated $B(E)$ distributions for
initial state binding energies of 0.2 MeV, 1.0 MeV and 5.0 MeV shown
in figure \ref{fig:beta2}. A fit with a single level did not produce
acceptable fits even though the channel radius was set to 4 fm. The
results for fits with two levels are displayed in table
\ref{tab:Rfit}, the agreement between calculated and fitted
distribution is perfect for the 5.0 MeV data, but there are systematic
deviations for the other two energies. The table gives the observed
position and full width at half maximum (in MeV) for the three $B(E)$
distributions as well as the R-matrix ``observed'' values
\cite{Lan58,Bay10} of the energies $E_i$, the widths $\Gamma_i$ and
the  beta strength parameters $B_i$ for transitions to the two levels.

\begin{table}
\centering
\caption{Results of R-matrix fits to beta
  strength distributions for potential depth 55.0 MeV. All energies
  are in units of MeV.}
\label{tab:Rfit} 
\begin{tabular}{ccccccccc}
 \hline
$E_i$ & $E_{max}$ & FWHM & $E_1$ & $\Gamma_1$ & $B_1$ & $E_2$ &
 $\Gamma_2$ & $B_2$ \\ \hline
0.2  & 0.41 & 0.30 & 0.45 & 0.38 & 0.44 & 360 & 8.0 & $4.8\cdot10^5$ \\
1.0  & 0.44 & 0.31 & 0.48 & 0.40 & 0.40 & 499 & $1.1\cdot10^4$ & 78 \\
5.0  & 0.45 & 0.33 & 0.49 & 0.41 & 0.31 & 39 & 1.21 & 282 \\ \hline
\end{tabular}
\end{table}

It is interesting that the differences in the maximum of the
distribution, $E_{max}$, are seen as well in the extracted fit values for $E_1$.
The shift of $E_{max}$ relative to 
$E_1$ is close to that given by equation (\ref{eq:eshift}). 
However, the shape of the distributions necessitated the
two component fits where in all cases the two R-matrix levels
interfere destructively between the two level positions. The width
parameters $\Gamma_1$ therefore do not correspond to the observed FWHM
values and the beta strength parameters $B_1$ are lower than the
maximum value of one. The parameters of the second level are poorly
constrained by the data and the fit uncertainties therefore
large. Nevertheless, the parameters $B_2$ are in all cases
unrealistically large, as is also the case for several values of $E_2$
and $\Gamma_2$. The overall fit can therefore not be interpreted in terms
of physically well-defined resonances,
it rather resembles the situation encountered in the R-matrix fits
\cite{Hyl10} to the decay of $^{12}$N into states in $^{12}$C above
the alpha particle threshold where unphysical values of $E$, $\Gamma$
and $B$ were obtained when a $0^+_4$ level was introduced in the fits
(this was done in order to reach an acceptable $\chi^2$-value).

Adding more levels in the R-matrix fit may in principle lead to a more
acceptable solution, but the extra parameters introduced as levels are
added implies that it will not be possible to determine all parameters
solely from the data. The contribution from the regions outside of the
nuclear core that occur in beta decay for small initial binding energy
can therefore be expected to give a difference in extracted resonance
energy between analyses of beta decay and elastic scattering data. In
our example this difference reached more than 10\% of the width of the
level.

\begin{figure}[thb]
\centering
     \includegraphics[width=8.cm,clip]{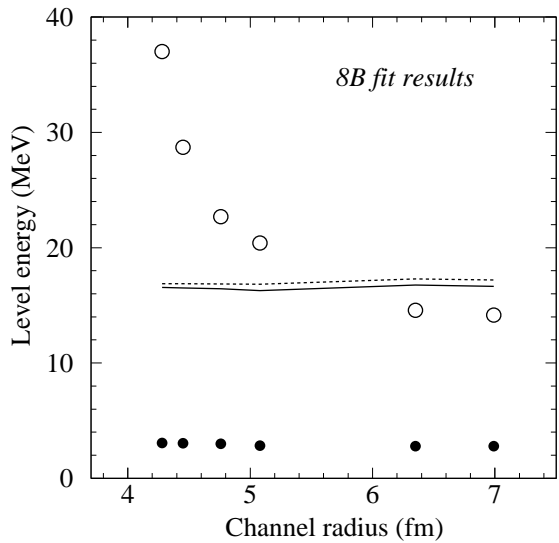} 
     \caption{The observed energies of the fitted levels from an
       R-matrix four-level fit \protect\cite{Hyl10a} of the beta-delayed alpha-spectrum from
       the decay of $^8$B is shown versus the channel radius used for
       the fit. The positions of the well established 3 MeV level
       (filled circle) and the $2^+$ doublet close to 17 MeV (lines) are
       stable whereas the position of the fourth level (open circle) varies.}
\label{fig:b8fit} 
\end{figure}

To learn more about how R-matrix fits may be interpreted we turn to the
example of the beta-delayed alpha decay of $^8$B which has been
studied on several occasions. We use the spectrum obtained recently in
\cite{Kir11} that also gives details on earlier work as well
as a brief account of the fitting procedure and results for a fit with
channel radius 4.5 fm. More details on the fit as well as results for
other channel radii can be found in \cite{Hyl10a}. To get an
acceptable fit one needs to include, apart from the well established
level at 3 MeV and the $2^+$ doublet at 16.6 MeV and 16.9 MeV, a
fourth $2^+$ level, see figure \ref{fig:b8fit} that shows the energy
positions as function of channel radius.  The nature of the extra
level has been discussed on several occasions, see e.g.\
\cite{War86,Bar89,Pag05,Bha06}, with Barker arguing from a
simultaneous fit to scattering and decay data that it is an intruder
state. However, not all R-matrix levels can be interpreted as
resonances \cite{Bre59,Lan58,Wig47}. For the case of $^8$Be a detailed
demonstration of this was made for the s-wave resonances in a
comparison \cite{Hum98} of R-matrix and K-matrix fits.

In more general terms this was demonstrated already by Wigner and
Eisenbud \cite{Wig47} who showed how R-matrix theory treats even the
case of vanishing interaction through an infinite number of levels $i$
with parameters $E_i = \hbar^2/(2M) (2i-1)^2 (\pi/(2a_c))^2$ and
$\gamma_i = (-1)^i \sqrt{\hbar^2/(Ma_c)}$. These levels cannot be
interpreted as resonances, but are needed in order to reproduce the
case of ``no scattering''. They do appear since the R-matrix levels
form a complete set of basis functions.  If, as is often the case, a
small number of R-matrix levels is used in a fit one may find the need
for a ``background level'' with unphysical energies or
$\gamma$-parameter, such a level effectively incorporates the effects
of the above infinite number of levels, as illustrated in
\cite{Hum98}. Turning this around, to incorporate continuum behaviour
in an R-matrix fit one needs extra levels with large $\gamma$ values
and approximate distance
\begin{equation}
   E_{i+1} - E_i = \frac{\hbar^2\pi^2}{Ma_c^2}  i \;.
\end{equation}
(The value for $a_c=$ 5 fm and $i=1$ is 8.25 MeV for the alpha-alpha
system and 18.0 MeV for the neutron system above.)
The fourth level in figure \ref{fig:b8fit} has a dependence on the
channel radius $a_c$ that is consistent with this behaviour and indeed
has quite large values of $\gamma$ (clearly larger than the ones for
the 3 MeV level, and therefore unphysical) and may therefore function as an
effective level. One would expect that introducing more levels in the
fits would give better behaviour, in particular at large $a_c$, but the
alternative approach that interprets the need for such extra levels as a
signature of a noticable direct contribution seems more attractive.
That fits performed at large channel radii may need more levels is
consistent with the general practice \cite{Bay10} of using rather
small values for the radii.

In summary, R-matrix theory has a complete basis and can reproduce any
decay mechanism, also decays directly to continuum states. The
presence of such decays can be indicated in several ways: by R-matrix
levels that do not correspond to any physical ``bump'' in the spectra,
by level energies that scale inversely as the channel radius squared,
or by levels that have unphysically large values for the parameters
that represent the coupling to different channels, like $\gamma$ and
$B$.

\section{Discussion}
\subsection{Other beta-delayed decays}
Most beta-delayed neutron emissions will not proceed through states
with single-particle neutron strength. The delta-shell potential
introduced above to mimic this effect may not be sufficiently realistic,
but the observed dependence on the initial state binding energy should
remain and could in extreme cases give peak shifts approaching the
width of the resonance.

The mirror process is proton emission following $\beta^+$ decay. The
Coulomb barrier will now affect both initial and final proton states
and diminish the contribution from distances beyond the nuclear
potential. On the other hand more decay channels will be open, there
is e.g.\ beta-feeding to the IAS for nuclei with more protons than
neutrons.  
A special situation may be encountered for isospin 1 systems
where (in light nuclei) the nuclei with isospin projection $\pm 1$ can
both decay to the same states in the nucleus with projection
$0$. These mirror decays may now differ in line shape.
Two possible cases occur for isospin 3/2 systems: 
The beta decay of $^{13}$O includes a transition to a level at 8.198
keV in $^{13}$N that is more than 200keV wide, this is the mirror
transition to the $^{13}$B case considered earlier. The proton energy
has only been measured in one experiment \cite{Knu05} where a surprising
energy shift 50 keV upwards with respect to reaction data was
reported. A more careful investigation of this case seems needed. 
The beta decay of $^{17}$Ne includes a transition to a quite broad,
700(250) keV, 8.2 MeV 3/2$^-$ level \cite{Bor88,Mor02} but the energy
of this resonance has not been measured precisely in any experiment.

The main remaining beta-delayed process to consider is beta-delayed
alpha decay (delayed emission of other particles, such as tritons, has
also been observed, but only in rather few nuclei). For such
transitions the overlap integrals will be more complex and simple
models hardly adequate. Nevertheless, there are, as alluded to above,
experimental data that may point to related effects. Consider first
the decays of $^8$B and $^8$Li into $2^+$ levels in $^8$Be that
subsequently decay by breaking up into two alpha particles.  As shown
in the previous section the most natural explanation of the $^8$B
decay data is in terms of a contribution also from direct decays to
continuum states. A detailed comparison of peak positions of the
lowest $2^+$ resonance in the two decays and in alpha-alpha scattering
performed in \cite{Bha06} gave the same value within about 20 keV, but
the beta strength is low in this case so the expected energy shift may
be small. Furthermore, the beta decay strength parameters for the two
mirror decays differed in the analysis and it may be worthwhile to try
to model the system rather than relying on R-matrix fits.

As the next example, consider decays leading to the alpha-particle
cluster states in $^{12}$C where the R-matrix fits mentioned above
\cite{Hyl10} also strongly indicated the presence of decays directly
to the continuum. Here one is dealing with a three alpha particle
final state which further complicates the theoretical
treatment. Existing cluster models \cite{Rot10,Kan07} may be a
starting point for an analysis, but in order also to evaluate the
overlap integrals that enter in the beta-decay process they have to be
extended. 

Finally, the E1 component of the astrophysically important
$^{12}$C($\alpha$,$\gamma$)$^{16}$O reaction is often evaluated by
combining reaction and beta decay data, see e.g.\ \cite{Buc09,Sch12}.
In the combined fit to data the resonance position and width is
assumed to be the same for all processes. In view of our results this
assumption should be checked.


\subsection{Other experimental probes}
It would be interesting to extend the calculations also to other ways
of populating resonances in order to see how large difference may be
there. A good starting point for transfer reactions could be the
semiclassical model for transfer to continuum states by Bonaccorso and
Brink \cite{Bon88}. Coupled channels calculations may also be used to
treat inelastic scattering. For knockout reactions at higher beam
energies there is already an extensive discussion on the reaction
mechanism, see e.g.\ \cite{Gar01} where arguments for a dominance of
direct transitions are presented.

It was striking to see the rather large difference in spectral shape
between elastic scattering and beta-decay that appeared for very wide
structures (to some extent even for s-wave systems), a difference that
must arise from the fact that the initial state in beta decay is
confined. This suggests that beta-decay is more sensitive to broad
resonances than elastic scattering. It will be interesting to see how
other experimental probes will behave and whether there will be
similar effects of spatially extended nuclear states (e.g.\ the rather
loosely bound deuteron often employed in transfer reactions). We have
so far only considered allowed beta-decay. In higher orders, $\lambda$,
one encounters matrix elements that, similar to the case of
electromagnetic interactions, involve spherical Bessel functions
$j_{\lambda}(qr)$ (or their generalization to take into account the
effects of the nuclear Coulomb field) with a wave number $q$ in the
relevant range for the transition \cite{Boh69}. This will lead to an
enhancement of direct transitions similar to that already mentioned
for electromagnetic transitions. The same conclusion also holds for
processes at somewhat higher energy, such as electron or neutrino
scattering or muon capture \cite{Zin06}, where the higher order
processes become more important.

\subsection{The non-resonant continuum ?}  \label{sec:nonres}
When calculating processes close to a threshold and employing a
framework, such as the Gamow Shell Model \cite{Dob07}, that is
sufficiently powerful to capture both bound state and continuum
behaviour, results are sometimes described in terms of resonant and
non-resonant continuum contributions. Such a division is of course, as
practitioners are well aware, only possible with a specific choice of
which resonances are included; the example of the calculation of the
$^{11}$Be dipole strength function \cite{Myo98} was quoted
already. A distinction may be made within a theoretical model, but
one cannot find a general and unambiguous experimental way of
separating resonant and non-resonant processes. This is well
established (see e.g.\ appendix 3F in \cite{Boh69}), but the above
calculations illustrate this in a new way.

The terms resonant and non-resonant continuum suggests that it is a
property of the continuum, i.e.\ of the final state on
its own. The cases discussed above where the initial state structure
influenced the detailed line shape shows that such an interpretation
can lead to a rather complex description of actual processes. In
a similar way radiative capture can in limiting cases be described as
proceeding through resonant capture or direct capture, but will in
general have contributions from both, and a general distinction
between one reaction mechanism and the other cannot be made. 
Assuming as in section \ref{sec:2res} that a resonance should be
localized in configuration space as well as in energy, one may use
short-distance and large-distance contributions as ways of
distinguishing resonant and continuum terms; see e.g.\ the explicit
discussion in \cite{Gar13} where cross terms also appeared, but note
that the exact point of division is arbitrary.

Based on the picture of spatial division one would distinguish
resonant and non-resonant contributions through their different radial
contribution. This explains why operators with different radial
weighting have different sensitivities to ``non-resonant''
contributions, with electromagnetic transitions being more prone to
display a direct mechanism. Except for situations where one of the two
contributions dominate, a clear-cut division is not possible.

\section{Conclusion}
The most important features of broad resonances and our main results
may be summarized as follows: 
\begin{itemize}
\item In the limit of broad resonances, different ways of defining a
  resonance will not have the same range of validity and will not give
  the same values for the resonance position and width.
\item Beta-decay (and to an even larger extent electromagnetic)
  transitions may get contributions from extra-nuclear distances, this
  distorts the observed line shape.  The effect can be naturally
  interpreted as a contribution from decays directly to continuum
  states. In general such contributions will increase in importance as
  one approaches the driplines,
  partly due to the low binding energies occuring there,
  partly due to the fact that more transitions with large beta
  strength will be seen. Transitions with large beta strength are more likely to show
  an effect.
\item Extraction of resonance parameters from data has to be done with
  care. The comparison of elastic scattering and beta-decay showed
  that R-matrix resonance parameters may depend on the
  process. R-matrix fits of continuum contributions require
  introduction of extra levels. Such levels often have unphysical
  parameter values: energies that depend on the channel radius or too
  strong coupling parameters.
\item Combining information from different types of experiments in
  order to narrow down the properties of a resonance can therefore go
  wrong for broad levels. A concrete example of this is the $2^+$
  levels in $^8$Be discussed in section \ref{sec:Rmat}.  Conversely,
  using resonance parameters from one experiment in another seemingly
  resonance dominated process may not be appropriate. In such
  cases a more involved theoretical analysis must be done.
\item Different experimental probes differ in their sensitivity to
  resonance structure, beta-decay may be more sensitive than other
  probes such as elastic scattering. This is due to the initial state
  in beta-decay being localized.
\item The levels that enter in R-matrix fits can not always be
  interpreted as resonances, the example discussed in detail involved
  the $^8$B decay.
\end{itemize}

Our results were obtained mainly for continuum p-wave neutrons, but
the more complex case of continuum s-waves was also considered
briefly. The trends found should persist for other beta-delayed
particle emission processes and related cases, and some of our
findings may also be relevant outside of nuclear physics for other
systems with short-range potentials situated close to a threshold.

\medskip
Acknowledgement.  
We would like to thank D.V. Fedorov and C.Aa. Diget for discussions.

\end{document}